\documentclass[singlespacing]{elsart}

\usepackage{graphicx}

\usepackage{amssymb}
\journal{Chemical Physics Letters}
\begin{document}

\begin{frontmatter}

\title{Exactly solvable Schr\"odinger equation with double-well potential for hydrogen bond}

\author{A.E. Sitnitsky},
\ead{sitnitsky@mail.kibb.knc.ru}

\address{Kazan Institute of Biochemistry and Biophysics, P.O.B. 30, Kazan
420111, Russia. Tel. 8-843-231-90-37. e-mail: sitnitsky@kibb.knc.ru }

\begin{abstract}
We construct a double-well potential for which the Schr\"odinger equation can be exactly solved via reducing to  the confluent Heun's one. Thus the wave function is expressed via the confluent Heun's function. The latter is tabulated in {\sl {Maple}} so that the obtained solution is easily treated. The potential is infinite at the boundaries of the final interval that makes it to be highly suitable for modeling hydrogen bonds (both ordinary and low-barrier ones). We exemplify theoretical results by detailed treating the hydrogen bond in  $KHCO_3$ and show their good agreement with literature experimental data.
\end{abstract}

\begin{keyword}
Schr\"odinger equation, confluent Heun's function, hydrogen bond.
\end{keyword}
\end{frontmatter}

\section{Introduction}
The search for a convenient analytically tractable double-well potential for the Schr\"odinger equation (SE) is a problem with long history (see e.g., \cite{Ros32}, \cite{Man35}, \cite{Raz80}, \cite{Kon86}, \cite{Kon95}, \cite{Xie12}, \cite{Dow13}, \cite{Che13}, \cite{Har14}, \cite{Tur10}, \cite{Tur16}, \cite{Dow16}, \cite{Dow17} and refs. therein). SE with a double-well potential is used in many well-known applications in physics and chemistry beginning from the inversion of the ammonia molecule and ending by heterostructures, Bose-Einstein condensates and superconducting circuits (see refs. in \cite{Xie12}, \cite{Dow13}, \cite{Che13}, \cite{Har14}). Besides it is a basic model for the description of ordinary hydrogen bonds (HB) and so-called low-barrier hydrogen bonds (LBHB) \cite{Mes02}, \cite{Mes03}, \cite{Mol06}, \cite{Mol07}. There are heated debates for more than twenty years that the latter may play an important role in biology (in particular in enzyme catalysis) \cite{Cle98}, \cite{Sho04}. The characteristic feature of LBHB is that a proton occupies a central position in the double well and the zero-point vibrational energy is close to the barrier top. Thus the commonly used semiclassical approximation (WKB) is principally inapplicable to the description of LBHB. The researches have to resort to numerical solutions of SE that greatly restricts the possibilities for scanning of the parameter space and makes the analysis to be cumbersome and inconvenient for perception. For this reason a relevant exact solution of SE would be highly useful for the analysis of LBHB nature.

The potentials from \cite{Ros32}, \cite{Man35}, \cite{Raz80}, \cite{Kon86}, \cite{Kon95} do not allow obtaining the natural eigenfunctions for the Schr\"odinger's operator. For this reason the authors expand the solution of SE over some suitable full sets of functions. The coefficients in the resulting series are determined from the appropriate recurrence relations. The latter makes the obtained solutions to be rather inconvenient for applications. The potentials from \cite{Xie12}, \cite{Dow13}, \cite{Che13}, \cite{Har14} are beneficially distinguished in this regard. For them the natural eigenfunctions for the Schr\"odinger's operator can be expressed via the confluent Heun's function (CHF) \cite{Ron95}, \cite{Sla00}. Thus the obtained solution of SE can be expanded in a series over natural eigenfunctions of Schr\"odinger's operator for the given potential. As a result the coefficients in the series become independent of one another. A lot of potentials for SE are shown to be solvable with the help of CHF \cite{Ish16}. The CHF is a known and by now well described special function which is a solution of the confluent Heun's equation \cite{Ron95}, \cite{Sla00}, \cite{Fiz12}, \cite{Fiz10}. In particular CHF is tabulated in {\sl {Maple}}. The latter makes its usage to be a routine procedure. This fact renders the obtained solution of SE very convenient for applications. Recently the exact solution of the Smoluchowski equation for reorientational motion in Maier-Saupe double well potential was obtained via CHF \cite{Sit15}, \cite{Sit16}. Here we apply similar approach to SE with a practically important type of double-well potential suitable for the description of hydrogen bonds.

Most of the potentials investigated with the help of CHF (\cite{Xie12}, \cite{Dow13}, \cite{Che13}, \cite{Dow16}, \cite{Dow17}) belong to the so-called quasi-exactly solvable ones for which only part of the spectrum can be computed exactly. The quantization is obtained from the requirement for the wave functions to contain the factors that are terminated polynomials. For the case of CHF it is reduced to a couple of the so-called polynomiality conditions \cite{Fiz12}, \cite{Fiz10}. One of them defines the quantization of the energy spectrum and the other sets the values of the parameters for the potential at which it is possible. Thus only at some peculiar values of the barrier height the exact energy levels and the corresponding wave functions can be computed. The potential investigated in \cite{Har14} is beneficially distinguished in the regard that it is exactly solvable. It enables one to determine the bound state energies for an arbitrary set of potential parameters. The boundary conditions are obtained from a very ingenious analysis of the wavefunction behavior at infinity along the line of \cite{Fer11}.

Unfortunately the  potentials studied in \cite{Xie12}, \cite{Dow13}, \cite{Che13}, \cite{Har14} are finite at the boundaries of infinite interval that makes them inapplicable to the investigation of HB. In the latter case the potential for the proton is naturally restricted by the location of heavy atoms (e.g., $O-H \cdot\cdot\cdot\cdot\cdot O$ or $N-H \cdot\cdot\cdot\cdot\cdot O$) and can be both symmetric or asymmetric even in at first sight symmetric cases like $O-H \cdot\cdot\cdot\cdot\cdot O$ (see e.g. \cite{Fil07}). For  LBHB the separation of heavy atoms is shorter and the bond is stronger (e.g., $N-H \bullet\bullet\bullet O$) \cite{Cle98}. The location of the heavy atoms makes the interval to be finite and the potential to be infinite at the boundaries. For this reason in the present paper we construct an exactly solvable potential for which the whole spectrum can be computed like for that from \cite{Har14} but which is infinite at the boundaries of the finite interval. The quantization results from the boundary conditions rather than from the requirement of terminated polynomials. The potential is infinite at the boundaries and thus is highly suitable for modeling HB and LBHB. We exemplify our solution by detailed calculation of the energy levels and corresponding proton wave functions for HB in $KHCO_3$ and show good agreement of theoretical results with literature experimental data.

The paper is organized as follows.  In Sec. 2 the problem under study is formulated.  In Sec. 3 the solution of SE is presented and the eigenvalues are found. In Sec. 4 the results for $KHCO_3$ are discussed and the conclusions are summarized. In Appendix some methodical comments are added.

\section{Schr\"odinger equation and the potential}
We consider the one-dimensional SE
\begin{equation}
\label{eq1} \frac{\hbar^2}{2m}\frac{d^2 \psi (x)}{dx^2}+\left[E-U(x)\right]\psi (x)=0
\end{equation}
where $U(x)$ is a double-well potential that is infinite at the boundaries of the finite interval $x=\pm L$.
We introduce the dimensionless energy $\epsilon$ and the dimensionless distance $y$
\begin{equation}
\label{eq2} \epsilon=\frac{8mL^2E}{\hbar^2 \pi^2}\ \ \ \ \ \ \ \ \ \ \ \ \ \ \ \ \ \ \ \ \ \ \ \ \ \ \ \ \ \ \ \ y=\frac{\pi x}{2L}
\end{equation}
so that $-\pi/2\leq y \leq \pi/2$. The dimensionless SE takes the form
\begin{equation}
\label{eq3}  \psi''_{yy} (y)+\left[\epsilon-U(y)\right]\psi (y)=0
\end{equation}
 For the potential we choose the model form
\begin{equation}
\label{eq4} U(y)=h\ \tan^2 y+c\frac{\sin y}{\cos^2 y}-b\sin^2 y+a\sin y
\end{equation}
Here $h$ and $c$ determine both the barrier width and the asymmetry, $b$ is the barrier height parameter and $a$ is the asymmetry parameter (along with $c$). An example of this potential (with particular choice $c=\sqrt {h}$) is depicted in Fig.1.
\begin{figure}
\begin{center}
\includegraphics* [width=\textwidth] {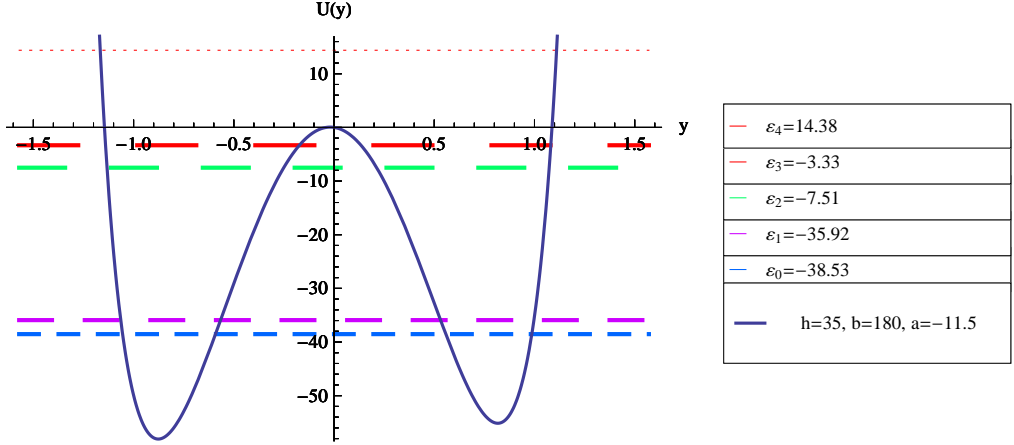}
\end{center}
\caption{The model double-well potential (\ref{eq21}) at the values of the parameters $h=35$, $b=180$, $a=-11.5$.  The parameters are chosen to describe the potential and the energy levels for the hydrogen bond in $KHCO_3$ crystal (experimental data are taken from \cite{Fil07}). The energy levels $\epsilon_0=-38.53$, $\epsilon_1=-35.92$, $\epsilon_2=-7.51$, $\epsilon_3=-3.33$ are respectively depicted by the dashes of increasing length. The dotted line is the first over barrier energy level from the infinite tier of them.} \label{Fig.1}
\end{figure}
The variation of $a$ practically does not touch upon the variation of the barrier width while the variation of $c$ alters the latter considerably.

\section{Reduction of Schr\"odinger equation to the confluent Heun's one}
CHF $\varphi (z) \equiv {\rm HeunC}\left(\alpha,\beta,\gamma, \delta,
\eta;z\right)$  (in the {\sl {Maple}} notation) is a solution of the confluent Heun's equation
\[
z(z-1)\varphi''_{zz}(z)+\left[\alpha z^2+(\beta+\gamma+2-\alpha)z-\left(\beta+1\right)\right]\varphi'_z(z)+
\]
\begin{equation}
\label{eq5}
\left[(\mu+\nu)z-\mu\right]\varphi (z)=0
\end{equation}
where
\begin{equation}
\label{eq6}\delta=\mu+\nu-\alpha\frac{\beta+\gamma+2}{2}
\end{equation}
\begin{equation}
\label{eq7}\eta=\frac{\alpha\left(\beta+1\right)}{2}-\mu-\frac{\beta+\gamma+\beta\gamma}{2}
\end{equation}
It should be mentioned that there is the second independent solution of (\ref{eq5})
but it is superfluous for the problem under consideration and is discarded further. Mathematically it means that in the general solution of (\ref{eq5})
\[
\varphi (z)^{general}=A\ {\rm HeunC}\left(\alpha,\beta,\gamma, \delta,
\eta;z\right)+B\ z^{-\beta}{\rm HeunC}\left(\alpha,-\beta,\gamma, \delta,
\eta;z\right)
\]
we set the constant $B$ to be zero $B=0$. Physically it stems from the fact that the second solution is
inconsistent with the requirement that for the infinite at the boundaries potential the wave function must be zero there.

We introduce a new variable $s$ by the relationship
\begin{equation}
\label{eq8} z=\frac{1+s}{2}
\end{equation}
The equation takes the form
\[
\left(1-s^2\right)\varphi''_{ss}(s)-\left[\frac{\alpha}{2} s^2+(\beta+\gamma+2)s+\left(\gamma-\beta-\frac{\alpha}{2}\right)\right]\varphi'_s(s)-
\]
\begin{equation}
\label{eq9} \left(\frac{\nu-\mu}{2}+\frac{\nu+\mu}{2}s\right)\varphi (s)=0
\end{equation}
We introduce a new variable $y$ by the relationship
\begin{equation}
\label{eq10} s=\sin y
\end{equation}
The equation takes the form
\[
\varphi''_{yy}(y)-\left[\frac{\alpha}{2} \sin^2 y +(\beta+\gamma+2)\sin y+\left(\gamma-\beta-\frac{\alpha}{2}\right)\right]\varphi'_y(y)\frac{1}{\cos y}-
\]
\begin{equation}
\label{eq11} \left(\frac{\nu-\mu}{2}+\frac{\nu+\mu}{2}\sin y\right)\varphi (y)=0
\end{equation}
Finally we introduce a new function  $\psi (y)$ by the relationship
\begin{equation}
\label{eq12} \psi(y)=\varphi (y)\exp \left(\frac{\alpha}{4}\sin y \right)\left(\cos y\right)^{\frac{\beta+\gamma+1}{2}}\left[ \tan\left(\frac{\pi}{4}+\frac{y}{2}\right)\right]^{\frac{\beta-\gamma}{2}}
\end{equation}
We set
\begin{equation}
\label{eq13} \gamma=-\left(\beta+1\right)
\end{equation}
and denote
\begin{equation}
\label{eq14} h=\left(\beta+\frac{1}{2}\right)^2
\end{equation}
\begin{equation}
\label{eq15} b=\frac{\alpha^2}{16}
\end{equation}
\begin{equation}
\label{eq16} c=\beta+\frac{1}{2}
\end{equation}
\begin{equation}
\label{eq17} a=\frac{\nu+\mu}{2}-\frac{\alpha}{4}
\end{equation}
\begin{equation}
\label{eq18} \epsilon=\frac{\mu-\nu}{2}-\frac{1}{4}\left(2\beta+1+\frac{\alpha}{2}\right)^2
\end{equation}
From (\ref{eq14}) and (\ref{eq15}) we express the parameters $\beta$ and $\alpha$ via those of the potential $h$ and $b$ but various choice of the signs is possible. The requirement of the agreement of the obtained theoretical results with literature experimental ones (see next Sec.) leads us to the correct choice
\begin{equation}
\label{eq19} \alpha=-4\sqrt {b}
\end{equation}
\begin{equation}
\label{eq20} \beta=-\frac{1}{2}+\sqrt {h}
\end{equation}
The other choice of the sign for $\beta$ is inconsistent with the requirement that for the infinite at the boundaries potential the wave function must be zero there. The other choice of the sign for $\alpha$ leads to an nonphysical solution that has no bound states.
As a result we obtain SE (\ref{eq3}) with the potential
\begin{equation}
\label{eq21} U(y)=h\ \tan^2 y+\frac{\sqrt {h}\ \sin y}{\cos^2 y}-b\sin^2 y+a\sin y
\end{equation}
whose solution is the wave function
\[
\psi(y)=\exp \left(-\sqrt {b}\sin y \right)\left[ \tan\left(\frac{\pi}{4}+\frac{y}{2}\right)\right]^{\sqrt {h}}{\rm HeunC} \Biggl(-4\sqrt {b},-\frac{1}{2}+\sqrt {h},
\]
\begin{equation}
\label{eq22} -\left(\frac{1}{2}+\sqrt {h}\right), 2a,
\frac{3}{8}-a-b-\frac{h}{2}-\epsilon;\frac{1+\sin y}{2}\Biggr)
\end{equation}
The potential (\ref{eq21}) is infinite at the boundaries. Thus the wave function must be zero there. This requirement at $y=-\pi/2$ is automatically satisfied due to $\tan$ in (\ref{eq22}). The requirement at $y=\pi/2$ can be satisfied only by a specific constraint imposed on ${\rm HeunC}$ and thus yields the boundary condition for determining the energy levels. As a result we obtain the equation for eigenvalues by setting $y=\pi/2$
\begin{equation}
\label{eq23}{\rm HeunC}\Biggl(-4\sqrt {b},-\frac{1}{2}+\sqrt {h},-\left(\frac{1}{2}+\sqrt {h}\right),2a,
\frac{3}{8}-a-b-\frac{h}{2}-\epsilon;1\Biggr)=0
\end{equation}
Its solutions form the spectrum of eigenvalues $\epsilon_n$ where $n=0,1,2, ...\ $ for the energy $\epsilon$.

It should be noted that in a few cases the solution of the confluent Heun's equation can be reduced to more simple functions (see \cite{Mai04}, \cite{Ish14}, \cite{Hou15}, \cite{Sha12} and refs. therein). However such cases impose severe constraints on the parameters of the potential. For instance the case 1. from \cite{Sha12} in our notation requires $a=\sqrt{b}$. The resulting potential with such asymmetry is very far from what we need for the description of hydrogen bonds. Other cases from \cite{Sha12} relate the parameters of the potential with the energy $\epsilon$ so that exact solutions can be obtained only for some specific energy levels. In our opinion a realistic asymmetric double-well potential can not be warped to satisfy above mentioned constraints.

\section{Results and discussion}
Fig.1 shows that the parameters of the potential (\ref{eq21}) can be chosen to provide precise description of the energy levels structure for a specific hydrogen bond. In Fig.1 the energy levels for the hydrogen bond in $KHCO_3$ crystal (experimental data are taken from \cite{Fil07}) are presented. The energy levels for the proton stretching mode along an ordinary hydrogen bond usually form a pair of doublets within the wells. The $O-H \cdot\cdot\cdot\cdot\cdot O$ in $KHCO_3$ crystal presents a typical example \cite{Fil07}. The ground-state splitting $\Delta E_{01}=E_1-E_0=216\ {\rm cm^{-1}}$ was determined from large scattering cross-section of protons. The transition frequencies for the upper states $\Delta E_{02}=E_2-E_0=2475\ {\rm cm^{-1}}$ and $\Delta E_{03}=E_3-E_0=2820\ {\rm cm^{-1}}$ were determined with help of the infrared and Raman spectroscopies. The distance between the minima of the potential $\approx 0.6\ \AA$ is known from the crystal structure. These experimental values are obtained from our dimensionless ones if we take $h=35$, $b=180$, $a=-8$ and set $L=0.68\ \AA$ (with taking into account that for a proton $m=1\ {\rm amu}$). The latter means that the distance between the minima is $\approx 0.65\ \AA$ that is in good agreement with the above mentioned estimate. The normalized wave functions are obtained from (\ref{eq22}) as follows
\begin{equation}
\label{eq24} \psi_i(y)^{normalized}=\frac{1}{Z_i}\psi_i(y)
\end{equation}
where
\begin{equation}
\label{eq25} Z_0=56;\ Z_1=-3338;\ Z_2=-41;\ Z_3=48
\end{equation}

They are depicted in Fig.2. The functions $\psi_0(y)^{normalized}$ and $\psi_1(y)^{normalized}$ correspond to the split ground state. The ground state splitting is rather large ($216\ {\rm cm^{-1}}$ in dimensional units). It results from the asymmetry of the potential. For the symmetric potential it would be smaller and the tunneling frequency would be $18\ {\rm cm^{-1}}$ \cite{Fil07}. Taking into account that $k_BT \approx 200\ {\rm cm^{-1}}$ at $T=300^{\circ} K$ one can conclude the functions describe the states with the population of the same order of magnitude (up to the factor $e^{-1}$) at room temperature. The highly delocalized functions $\psi_2(y)^{normalized}$ and $\psi_3(y)^{normalized}$ describe excited states. It should be stressed that the latter are rather close to the barrier top (see Fig.1). Thus the functions $\psi_2(y)^{normalized}$ and $\psi_3(y)^{normalized}$ can not be described in semiclassical approximation (WKB). The exact analytic representation of these wave functions is the merit of the present approach.

\begin{figure}
\begin{center}
\includegraphics* [height=5cm] {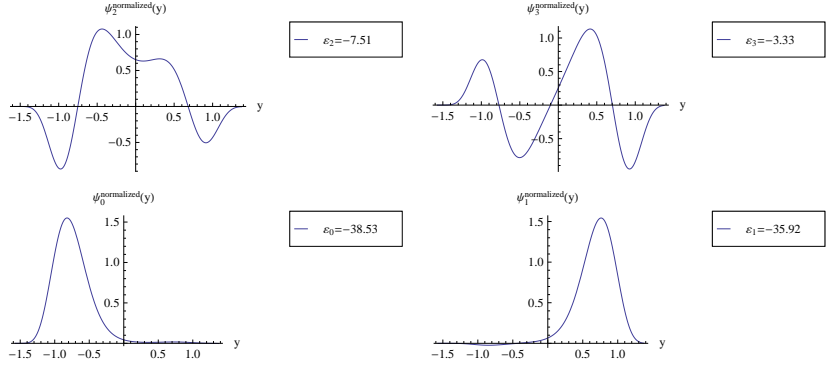}
\end{center}
\caption{Normalized wave functions (\ref{eq24}) for Schr\"odinger equation (\ref{eq3}) with the double-well potential (\ref{eq21}) corresponding to the energy levels $\epsilon_0=-38.53$, $\epsilon_1=-35.92$, $\epsilon_2=-7.51$, $\epsilon_3=-3.33$ for the hydrogen bond in $KHCO_3$ crystal (experimental data are taken from \cite{Fil07}).} \label{Fig.2}
\end{figure}

The potential in Fig.1 is slightly asymmetric (the left well is deeper than the right one). The wave functions are largely (but not entirely) localized in the left and right wells respectively. It is in accordance with the qualitative fact that a small fraction of the wave functions is delocalized in the other well \cite{Fil07}. The asymmetry of the potential leads to tiny delocalization of the probability density for the corresponding wave functions and is known to play a key role in spectroscopy and proton-transfer dynamics \cite{Fil07}. We obtain excellent agreement of our wave functions with those calculated numerically in \cite{Fil07} for a suitable double-well potential. However it should be stressed that although the numerical analysis of SE with a double-well potential can yield identical results with those of the analytic treatment the latter has a substantial advantage. The parameter space of a double-well potential is extremely vast and to find their necessary combination to describe experimental data for the energy levels structure of a specific hydrogen bond with the help of the numerical solution of SE seems to be a daunting task. In our case the solution of the obtained equation (\ref{eq23}) with the help of {\sl {Maple}} software package facilitates the problem enormously. The latter strategy provides an efficient tool for scanning of the parameter space of the model. Another advantage of our analytic solution over the numerical one is the flexibility in the implementation of subsequent analysis (calculation of transition rates for proton-transfer dynamics, their corrections due to tunneling effect, etc.). We do not touch upon this matter in the present article and leave it for further work. Nevertheless it is obvious that the availability of the exact analytic expression for the wave function enables one much wider possibilities in treating the system under investigation than purely numerical solution of SE.

One can conclude that the suggested double-well potential for the Schr\"odinger equation is exactly solvable. The analytic expressions for the wave functions are obtained via the confluent Heun's one. The equation for the eigenvalues (i.e., energy levels) can be easily solved numerically with the help of {\sl {Maple}} software package. Besides the potential is infinite at the boundaries of the final interval that makes it to be highly suitable for modeling hydrogen bonds (both ordinary and low-barrier ones). The parameter space for the potential is rich enough to provide precise quantitative description of relevant experimental data on hydrogen bonds.

\section{Appendix}
Although {\sl {Maple}} treats (\ref{eq23}) very efficiently it proves to be too time consuming in evaluating integrals with CHF, e.g. those needed for the normalization of the wave functions. The author of the present article can not explain the inefficiency of $int(...,numeric)$ in this case and refer the reader to \cite{Fiz12} where some drawbacks of ${\rm HeunC}$ realization in {\sl {Maple}} are expertly discussed.
In contrast with the help of the series representation of ${\rm HeunC}$ (with large but finite number of terms)
one can obtain a very efficient tool for evaluating the corresponding integrals.
The series expansion of ${\rm HeunC}$ \cite{Ron95}, \cite{Sla00}, \cite{Fiz10} in our case takes the form
\[
{\rm HeunC} \Biggl(-4\sqrt {b},-\frac{1}{2}+\sqrt {h},-\left(\frac{1}{2}+\sqrt {h}\right),2a,
\frac{3}{8}-a-b-\frac{h}{2}-\epsilon;\frac{1+\sin y}{2}\Biggr)=
\]
\begin{equation}
\label{eq26}
\sum_{k=0}^{\infty}v_k \left(\frac{1+\sin y}{2}\right)^k
\end{equation}
The coefficient $v_k$ is determined from the three-term recurrence relation
\begin{equation}
\label{eq27} A_k v_k=B_k v_{k-1}+C_k v_{k-2}
\end{equation}
with the initial conditions $v_{-1}=0$, $v_0=1$. Here
\begin{equation}
\label{eq28} A_k=1+\frac{\sqrt{h}-1/2}{k}
\end{equation}
\begin{equation}
\label{eq29} B_k=1+\frac{4\sqrt {b}-2}{k}+\frac{1-a-b-h-\epsilon-3\sqrt{b}+2\sqrt{bh}}{k^2}
\end{equation}
\begin{equation}
\label{eq30} C_k=\frac{4\sqrt{b}}{k^2}\left(\frac{a}{2\sqrt{b}}+\frac{3}{2}-k\right)
\end{equation}
We find that the visual identity of the series representation of ${\rm HeunC}$ with that given by {\sl {Maple}} can be attained if we take some large but finite number $N$ for terms in the series instead of the infinity in (\ref{eq26}). $N$ can be small for $\psi_0(y)$ but increases with the number $i$ in $\psi_i(y)$ and for
$\psi_4(y)$ it should be $N>230$. Having made sure that both representations of ${\rm HeunC}$ are visually identical we (making use of the series representation of ${\rm HeunC}$) perform the numerical calculations of the required integrals by standard methods. If one avoids the enclosed cycles in calling the three-term recurrence relation then the calculation proceeds within seconds on any PC of even low power.

Acknowledgements. The author is grateful to Dr. Yu.F. Zuev
for helpful discussions. The work was supported by the grant from RFBR N 15-44-02230.


\newpage
\begin{thebibliography}{00}
\bibitem{Ros32}
N. Rosen, P.M. Morse, On the vibration of polyatomic molecules, Phys. Rev. 42 (1932) 210-217.
\bibitem{Man35}
M.F. Manning, Energy levels of a symmetrical double minima problem with applications to the $NH_3$ and $ND_3$ molecules, J.Chem.Phys. 3 (1935) 136-138.
\bibitem{Raz80}
M. Razavy, An exactly soluble Schrodinger equation with a bistable potential, Am. J. Physics,  48 (1980) 285-288.
\bibitem{Kon86}
H. Konwent, One-dimensional Schr\"odinger equation with a new type double-well potential, Physics Letters, A118 (1986) 467-470.
\bibitem{Kon95}
H. Konwent, P. Machnikowski, A. Radosz, A certain double-well potential related to SU(2) symetry, J. Phys. A: Math. Gen. 28 (1995) 3757-3762.
\bibitem{Xie12}
Qiong-Tao Xie, New quasi-exactly solvable double-well potentials, J. Phys. A: Math. Theor. 45 (2012) 175302.
\bibitem{Dow13}
C. A. Downing, On a solution of the Schr\"odinger equation with a hyperbolic double-well potential, J. Math. Phys. 54 (2013) 072101.
\bibitem{Che13}
Bei-Hua Chen, Yan Wu, Qiong-Tao Xie, Heun functions and quasi-exactly solvable double-well
potentials, J. Phys. A: Math. Theor. 46 (2013) 035301.
\bibitem{Har14}
R.R. Hartmann, Bound states in a hyperbolic asymmetric double-well, J.Math.Phys. 55 (2014) 012105.
\bibitem{Tur10}
A.V. Turbiner, Double Well Potential: Perturbation Theory, Tunneling, WKB (beyond instantons),
Intern.Journ.Mod.Phys. A 25 (2010) 647-658.
\bibitem{Tur16}
A.V. Turbiner, One-dimensional quasi-exactly solvable Schrodinger equations, Physics Reports 642 (2016) 1-71.
\bibitem{Dow16}
C.A. Downing, M.E. Portnoi, Magnetic quantum dots and rings in two dimensions, Phys.Rew. B 94 (2016) 045430.
\bibitem{Dow17}
C. A. Downing, Two-electron atom with a screened interaction, Phys. Rev. A 95 (2017) 022105.
\bibitem{Mes02}
D.A. MacDonald, G.E. Eppard, C.J. Halkides, M. Messina, A critical comparison of approximation methods and models for equilibrium properties of low-barrier hydrogen bonds, J.Chem.Inf.Comput.Sci. 42 (2002) 1390-1397.
\bibitem{Mes03}
L.A. Butler, J.E. Miller, C.J. Halkides, M. Messina, On the possibility of using UV spectroscopy as a measure
of the low-barrier hydrogen bond, Structural Chemistry 14 (2003) 605-616.
\bibitem{Mol06}
M.A. Porter, P.A. Molina, The low-barrier double-well potential of the $O^{\delta 1}-H-O^{\delta 1}$
hydrogen bond in unbound HIV protease: A QM/MM characterization, J.Chem.Theory Comput.  2 (2006) 1675-1684.
\bibitem{Mol07}
R.N. Karingithi, C.L. Shaw, E.W. Roberts, P.A. Molina, The probability distribution function as a descriptor of hydrogen bond strength, Journal of Molecular Structure: THEOCHEM  851 (2008) 92-99.
\bibitem{Cle98}
W.W. Cleland, P.A. Frey, J.A. Gerlt, The low barrier hydrogen bond in enzymatic catalysis, J.Biol.Chem.  273 (1998) 25529-25532.
\bibitem{Sho04}
M. Shokhen, A. Albeck, Is there a weak H-bond $\rightarrow$ LBHB transition on tetrahedral
complex formation in serine proteases? PROTEINS: Structure, Function, and Bioinformatics  54 (2004) 468-477.
\bibitem{Ron95}
A. Ronveaux (ed.), Heun Equations, Oxford Univ. Press, 1995.
\bibitem{Sla00}
S.Y. Slavyanov, W. Lay, Special functions, a unified theory based on singularities, Oxford: Oxford Mathematical Monographs, 2000.
\bibitem{Ish16}
A. Ishkhanyan, Schrodinger potentials solvable in terms of the confluent Heun functions, Theoretical and Mathematical Physics 188 (2016) 980-993.
\bibitem{Fiz12}
P.P. Fiziev, D.R. Staicova, Solving systems of transendental equations involving the Heun functions,
AJCM 2 (2012) 95-105.
\bibitem{Fiz10}
P.P. Fiziev, Novel relations and new properties of confluent Heun functions and
their derivatives of arbitrary order, J. Phys. A: Math. Theor. 43 (2010) 035203.
\bibitem{Sit15}
A.E. Sitnitsky, Exact solution of Smoluchowski's equation for reorientational motion in Maier-Saupe potential,
Physica A: Statistical Mechanics and its Applications 419 (2015) 373-384.
\bibitem{Sit16}
A.E. Sitnitsky, Probability distribution function for reorientations in Maier-Saupe potential,
Physica A: Statistical Mechanics and its Applications 452 (2016) 220-228.
\bibitem{Fer11}
F.M. Fern\'andez, Wronskian method for bound states, Eur. J. Phys. 32 (2011) 723-732.
\bibitem{Fil07}
F.Fillaux, A. Cousson, M.J.Gutmann, Proton transfer across hydrogen bonds: From reaction path to Schr\"odinger's cat, Pure Appl.Chem. 79 (2007) 10223-1039.
\bibitem{Mai04}
R.S. Maier, On reducing the Heun equation to the hypergeometric equation, J. Differential Equations 213 (2005) 171-203.
\bibitem{Ish14}
T.A. Ishkhanyan, A.M. Ishkhanyan, Expansions of the solutions to the confluent Heun equation
in terms of the Kummer confluent hypergeometric functions, AIP ADVANCES 4 (2014) 087132.
\bibitem{Hou15}
M.N. Hounkonnou, A. Ronveaux, On some differential transformations of hypergeometric equations, Journal of Physics: Conference Series 597 (2015) 012044.
\bibitem{Sha12}
V.A. Shahnazaryan, T.A. Ishkhanyan, T.A. Shahverdyan, A.M. Ishkhanyan, New relations for the derivative of the confluent Heun function, Armenian Journal of Physics 5 (2012) 146-155.
\end{thebibliography}
\end{document}